\documentclass[aps,prr,twocolumn,superscriptaddress,longbibliography]{revtex4-1}

\usepackage{color}
\usepackage{epsfig}
\usepackage{amsmath}
\usepackage{amssymb}
\usepackage[normalem]{ulem}
\newcommand{\myeq}[2][]{Eq#1.~(\ref{#2})}
\newcommand{\myfig}[2][]{Fig#1.~\ref{#2}}

\begin{document}

\title{Externally driven local colloidal ordering induced by
  a point-like heat source}

\author{Nicolas Bruot}
\email[Email: ]{nicolas@bruot.org}
\affiliation{Institute of Industrial Science, The University of Tokyo,
  4-6-1 Komaba, 153-8505 Meguro-ku, Tokyo, Japan}
\affiliation{Institut de Physique de Nice, CNRS UMR 7010, Universit\'e
  C\^ote d'Azur, 28 avenue Joseph Vallot, 06108 Nice, France}
\author{Hajime Tanaka}
\email[Email: ]{tanaka@iis.u-tokyo.ac.jp}
\affiliation{Institute of Industrial Science, The University of Tokyo,
  4-6-1 Komaba, 153-8505 Meguro-ku, Tokyo, Japan}

\begin{abstract}
  We study here how phase transitions are induced in colloidal
  suspensions by a point-like heat source, an optically trapped metal
  oxide particle absorbing light. We find that thermophoresis
  increases the number density of colloids around the oxide particle,
  leading to the appearance of solid clusters.  Our analysis based on
  thermophoresis reveals that the solid-fluid interface position is
  purely determined by the relationship of the particle concentration
  profile in the fluid state with the volume fraction of the phase
  transition, and no other effect of thermodynamics is seen in the
  cluster sizes. In this system, we observe the formation of
  face-centered cubic crystals, amorphous states, and structures with
  icosahedral order. This shows a rich possibility of non-trivial
  orderings under spatially controlled heterogeneous growth in
  external ``semi-soft'' potentials that are softer than walls but
  with substantial variations at the scale of a few particle
  diameters. Because of the tuneable rate of addition of particles to
  the clusters, we propose this method could be used to study the
  complex interplay of densification with local spatial confinement
  effects, kinetics, and ordering.
\end{abstract}

\maketitle

\section{Introduction}

Colloidal particles dispersed in a solvent are excellent systems to
investigate phase transition phenomena as the particles experience
thermal fluctuations and at the same time they are big enough to be
observed at the particle-level resolution with common microscopy
techniques and at relevant timescales~\cite{lowen13}.  In particular,
confocal microscopy observations provide microscopic information like
individual particle locations and dynamics, which has tremendously
contributed to the fundamental understanding of equilibrium and
non-equilibrium phase transitions.  They include various phenomena
such as crystal nucleation~\cite{gasser01,tan13}, glass
transition~\cite{kasper98,kegel00,weeks00,weeks02,schall07,leocmach13,hallett18,pinchaipat17},
phase separation and
gelation~\cite{lu2008gelation,royall13,arai17,tsurusawa2019direct}.

Since phase transitions of colloids can be affected by external
fields, we may induce various types of transitions by applying
external fields in a reversible fashion, such as
optical~\cite{loudiyi92,wei98,verhoeff15},
magnetic~\cite{zahn99,zhao13,van_blaaderen13},
electric~\cite{zhang04,zhao13,van_blaaderen13} or thermal fields
(taking advantage of either
thermophoresis~\cite{moore10,jiang09,deng12} or temperature-dependant
particle size~\cite{peng15,wang15}).
Among these, the optical field is particularly interesting since it
can be spatially localized in a region less than the size of
colloids. In particular, local heating due to light absorption
provides a convenient way to induce {\it localized phase transitions
  under open out-of-equilibrium}
conditions~\cite{piazza2008thermophoresis,wurger2010thermal}. Local
heating may perturb colloidal suspensions in a variety of ways.  It
should lead to more vigorous Brownian motion in the heated
region. Furthermore, a coupling to colloidal particles with a
temperature gradient, termed as the thermophoretic effect, leads to
migration or depletion, i.e., the local change in the volume fraction
of colloids $\phi$ in a heated
region~\cite{piazza08,piazza2008thermophoresis}. Local heating may
also induce convective flows if it is strong
enough~\cite{duhr05}. Since many colloid phase transitions are induced
by the change in $\phi$, we may induce phase transitions by spatially
modulating $\phi$ via such a coupling.  It opens up a new type of
localized phase transitions under energy input, applicable to
gas-liquid transitions, crystallization, glass transition, phase
separation and gelation.  For example, Jiang \emph{et
  al.}~\cite{jiang09} found that migration of colloids can be driven
by a nonuniform polymer distribution sustained by the polymer's
thermophoresis due to laser heating.  These studies may serve as model
cases of phase transitions taking place in nature under a variety of
inhomogeneous fields in a complex manner.

In this Letter, we study local heating effects on phase transitions
and packing of the simplest possible colloidal system commonly used to
study phase transitions: hard spheres.  We create a very localized
inhomogeneous temperature field by shining a laser on a titanium
dioxide particle immersed in a colloidal suspension. This oxide sphere
(impurity) is both trapped and heated by optical tweezers, inducing
strong thermophoretic effects, which move the particles of the
suspension towards the hot oxide particle.  The direction of particle
motion is opposite to that expected for uncharged
particles~\cite{wurger09}, indicating crucial roles of surface charges
and/or chemistry, and solvent properties.  As a consequence of
inhomogeneous densification, we observe unconventional nucleation and
growth of condensed solid phases and other states (crystals,
structures with icosahedral order, and amorphous solids) under the
inhomogeneous temperature field.  The steady-state size and dynamics
of growth are investigated.

\begin{figure}
  \centering
    \includegraphics{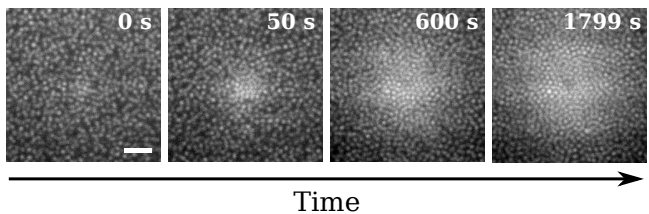}
    \caption{\label{fig:time_seq} Confocal imaging snapshots showing
      the growth of a polycrystalline structure (particles radius: $a
      = 0.40~\mu$m).  At $t = 0$~s, the laser power is switched from
      0.2 to 0.7~W in a sample of bulk volume fraction $\phi_\infty =
      0.25$ (scale bar: $5~\mu$m).  The heated titania particle at the
      center is barely visible as it slightly out of the focal plane
      because of gravity.}
\end{figure}

\begin{figure}[h!]
  \centering
    \includegraphics[width=8.5cm]{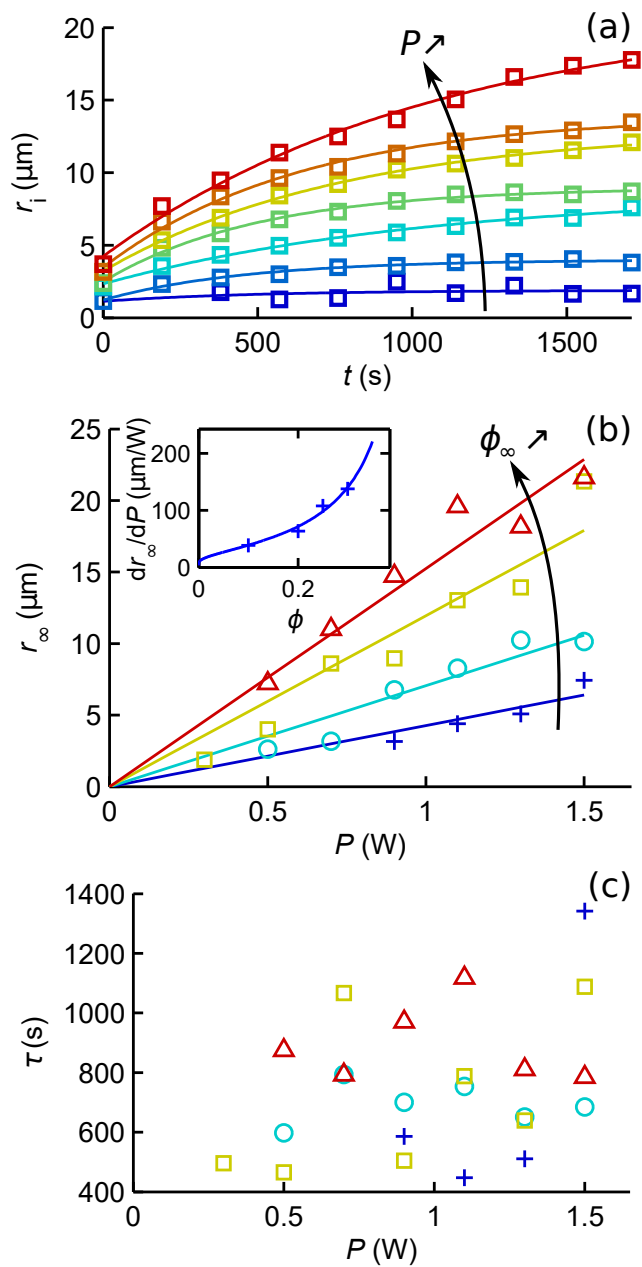}
    \caption{\label{fig:sizes_graphs} Size measurements of the
      condensed part of the aggregates for PMMA colloids of radius $a
      = 0.40~\mu$m.  (a) Size growth as a function of time for
      $\phi_\infty = 0.25$ and $P = 0.3$, 0.5, 0.7, 0.9, 1.1, 1.3 and
      1.5~W (from blue to red).  The experiments (markers) are well
      fitted by exponential convergences (lines).  (b) Final size
      $r_\infty$ of the polycristalline clusters depending on laser
      power and volume fraction (markers: $\phi_\infty =$~0.1 (+), 0.2
      ($\circ$), 0.25 ($\square$) and 0.3 ($\bigtriangleup$)).  For
      each volume fraction, the lines represent proportional fits.
      The slope of these fits is shown in the inset, depending on
      volume fraction (markers).  The data are fitted to the model
      described by \myeq{eq:r_i_formula} (line).  (c) Characteristic
      growth time $\tau$ depending on laser power and volume fraction
      (same coloring as in (b)).}
\end{figure}

\section{Materials and methods}

The samples are nearly the same as those described in
Refs.~\cite{arai17,tsurusawa2019direct} and we only describe here
their main features and the novelties.  Poly(methyl methacrylate)
(PMMA) colloids stabilized with metacryloxypropyl-terminated
poly(dimethylsiloxane) and containing a small amount of Rhodamine-B
dye were synthesized.  We always used organic solvents that were kept
dry by storing them in contact with molecular sieves (Sigma-Aldrich,
rods, 3A, 1/8). We suspended the spherical colloids of radii varying
from $a = 0.15$ to $1.2~\mu$m in a solution of about 20~\% of
\emph{cis}-decahydronaphtalene (\emph{cis}-decalin, $>98~\%$, TCI) and
bromocyclohexane (98~\%, Aldrich) saturated in
tetra-\emph{n}-butylammonium bromide salt (Wako).  Such samples are
known to have interactions and phase behavior close to hard
spheres~\cite{leocmach13,arai17,royall13,tsurusawa2019direct}.
Samples with different volume fractions between 5~\% and 30~\%, far
from the freezing point ($\sim 49.4$~\%), were prepared.  In these
mixtures, the particles are refractive index-matched, and the exact
ratio of the solvents is tuned to have a good density match that
prevents the sedimentation of the particles for 10 hours at least.
Separately, titania metal oxide particles of radius $R_0 = 1.5~\mu$m,
which were initially in an aqueous solution (Corpuscular, -COOH
functionalized), are transferred to \emph{cis}-decalin using ethanol
($99.5~\%$, Wako) as an intermediate solvent.  2$~\mu$L of this
solution is added to 200$~\mu$L of the solution of PMMA particles and
stirred.  Such solutions of PMMA colloids and titania particles were
sealed in glass capillary tubes suitable for optical trapping and
confocal microscopy observation (VitroTubes, 100~$\mu$m $\times 2$~mm
$\times$ 50~mm).
We found that it is necessary to coat the walls of the samples to
prevent the titania particles from sticking to the glass.  Therefore,
the inside of the capillaries with two parallel walls spaced by
100~$\mu$m was covered with a thin layer of PMMA by flowing a solution
of PMMA dissolved in toluene, and then flowing nitrogen to spread the
solution on the inner glass surfaces and to remove the excess solution
and evaporate the toluene.  This resulted in a thin solid PMMA film of
thickness $\lesssim 1~\mu$m on all the inner surfaces of the sample,
which was checked by confocal microscopy in test samples using a
fluorescent PMMA/toluene solution.  The capillaries were finally
filled with a solution of PMMA colloids and titania particles, and
then put on a microscope slide and sealed with UV-cured glue (NOA 68).
The samples were not filled entirely so that air initially separated
the glue and the solution.  The presence of air provided three
advantages.  Firstly, the liquid glue is never in contact with the
solution, which limits the possibility of compounds from the glue
being dissolved in the solution before it is cured.  Secondly, the
center of the sample containing the solution could easily be protected
from UV light during curing, which was necessary as we found that long
direct UV illumination led to particles sticking to each other and
forming large solid three-dimensional structures.  Thirdly, the
solution in the finished capillaries could be stirred by shaking the
sample vigorously to make air bubbles travel several times from one
side of the sample to the other.  Such shaking was done at the
beginning of each day of the experiment to homogenize the samples that
would otherwise show significant sedimentation over several days.

Samples were placed on a setup consisting of an Olympus IX81
microscope with Olympus UPlanSApo 60x, 1.20 NA water immersion
objective.  Observation of the PMMA particles was done by confocal
microscopy using a 532~nm excitation laser (JUNO Nd:YVO4, 500~mW), a
Yokogawa CSU-X1 confocal unit and an EMCCD Andor camera (iXon
DV885KCS-VP).
Optical tweezing was obtained using a holographic optical tweezers
(HOTkit, Arryx) setup combined to the microscope, although we did not
use in this work the holographic features.  In this configuration, the
output of a 1064~nm-wavelength fibred laser (IPG, YLR-10-LP, 10~W)
passes successively through a half-wave plate and a 50:50 beamsplitter
cube, is reflected on a spatial light modulator (SLM, 512x512, Boulder
Nonlinear Systems) that is here turned off and thus essentially used
as a mirror, and is directed back on the beamsplitter, after which it
passes through a telescope.  In the holographic configuration, the
telescope contains a mask at the focal point of the lenses to block
the 0th order of the beam diffracted by the SLM.  This mask was
removed here so that the main beam could be used for trapping.  After
the telescope, the beam has a Gaussian shape with a size slightly
greater than the back aperture of the objective to which it is sent.
The beam is directed to the back port of the microscope and reflects
inside the microscope on a dichroic mirror (950~nm shortpass, Semrock
FF01-950/SP) before finally entering the objective.

Before the start of an experiment, a titania particle was trapped
close to the waist of the laser beam and the sample was moved to have
the particle at about a 50~$\mu$m distance from the bottom and top
walls of the cell.  During this stage, a low laser power of 0.2~W was
used.  Powers indicate values from the laser controller. According to
our estimate, for 1~W on the controller, the actual power passing
through the sample is of less than 235~mW, as this value was measured
at a location before entering the objective.

\section{Results}

\subsection{Measurements of aggregates sizes}

When switching the laser power from a low value (0.2~W) to a higher
value $P$ (up to 1.5~W), aggregates start growing for sufficiently
high $P$, until a steady state is reached.  \myfig{fig:time_seq} shows
confocal images of the plane containing the laser focal point at
different times for an experiment at the sample volume fraction
$\phi_\infty = 0.25$, $P = 0.7$~W and $a = 0.4~\mu$m.  Supplemental
movies S1$\sim$S3~\cite{suppl_movies} also show the growth,
accelerated 100 times, for $\phi_\infty = 0.1$ and $P = 0.3, 0.7$ and
$1.1$~W, respectively.  Clearly ordered and large polycrystalline,
structures are present for a long time, showing that a phase
transition occurred. We can also see that the nearby fluid phase has a
higher concentration of particles than the bulk of the fluid.  In
other samples with large PMMA particles of radius of $a=1.2$~$\mu$m,
we could identify the existence of face-centered cubic (fcc)
structures in the solid part of the aggregates.
The aggregation does not occur when the laser is shinned in the sample
without a titania particle: for small particles of radius $a =
0.15~\mu$m, the laser is not able to trap the PMMA colloids while for
a radius of 1.2~$\mu$m, PMMA particles could be trapped occasionally
because of an imperfect refractive-index match, but no other particles
aggregated around it.

From movies consisting of time series of images as in
\myfig{fig:time_seq}, we have extracted the size of the solid
condensed structures for colloids of radius $a = 0.40~\mu$m.  Because
of the small size of particles, the high density of the aggregates and
the limitation of the recordings to a two-dimensional plane, particle
tracking could not be performed accurately.  We instead measure a size
based on a locally averaged quantity related to the mobility of the
particles, which is described in Appendix~\ref{app:size_meas}. We
assume solid clusters have a spherical shape and thus characterize the
size simply by the cluster radius $r_{\text i}$.
\myfig{fig:sizes_graphs}(a) shows growth profiles for bulk volume
fraction $\phi_\infty = 0.25$ when setting at $t = 0$~s the laser
power from 0.2~W (for which there are no aggregates) to various values
$P$.  The size increases faster with higher laser power.  It also
reaches higher steady-state values, as can be seen at least on the low
$P$ curves.  This size increase with time is well fitted empirically
to exponential approaches to a finite radius: $r_{\text i}(t) =
r_\infty + (r_0 - r_\infty) \mathrm{e}^{-t/\tau}$, thus giving
estimates of the final size $r_\infty$ of the aggregates and the
characteristic timescale of the growth of $\tau$.  These two
quantities are shown in Figs.~\ref{fig:sizes_graphs}(b) and (c) as
functions of $\phi_\infty$ and $P$.  In \myfig{fig:sizes_graphs}(b),
the final size grows linearly with laser power, and the dependency
with the volume fraction is super-linear, as is evidenced by the
linear fit for each volume fraction whose slope is plotted in the
inset.
The timescale of growth $\tau$ shows very scattered data with no clear
dependency on either $P$ or $\phi_\infty$.

\subsection{Origin of the formation of aggregates}

In the experiments, aggregates formed only in the presence of the
trapped central oxide particle.  Their size can be much greater than
the spatial window in which optical forces could act on the
surrounding colloid particles, since the spot size estimated from the
Rayleigh criterion is $1.2 \lambda / NA \approx 1~\mu$m, assuming a
uniformly lit objective.  Hence, we can rule out the possibility that
the aggregation emerges from optical forces on the colloidal
particles.  Instead, we show that the particles are subject to
thermophoresis, due to the heating of the trapped titania particle by
the tweezers laser.

We have tested that PMMA particles move towards hot regions in a
temperature gradient, that is, by thermophoresis with a negative Soret
coefficient.  In a sample of PMMA colloids (without titania particles)
containing a platinum wire of diameter 60~$\mu$m, we flow an electric
current through the wire to heat the fluid locally and create a
temperature gradient.  Confocal movies imaging planes cutting through
a diameter of the wire (movie S4) and a cross-section (movie S5) were
taken.  Shortly after the start of the movies that are accelerated 20
times, a current of 50~mA is applied to the wire, which is turned off
after several minutes.  The times at which the current is turned on
and off are evidenced by the wire moving to the left or right because
of dilation.  These movies show that particles aggregate around the
wire and that no convective flows arise (in movie S4, the gravitation
field is vertical).  Moreover, cutting the current leads to the
aggregates disappearing within a few minutes.  These tests confirm
that the aggregation originates from thermophoresis and not from
optical forces in the experiments with the trapped titania
particles. They also rule out any photo-induced phenomenon as seen by
Kim, Shah and Solomon~\cite{kim14a}, which is further supported by the
fact that we do not see any strong dependency of $\tau$ with
$\phi_\infty$ as opposed to their experiments.

\subsection{Model for the steady-state size}

To support further that the thermal gradient is the cause of the
aggregation, we show below that the experiments in
\myfig{fig:sizes_graphs} are consistent with a model based on
thermophoresis.  We assume that significantly more laser light is
absorbed by the titania particle than the fluid or the PMMA particles,
as we see no aggregates when no titania particle is trapped.
Therefore, the central metal oxide particle acts as a small,
point-like, heat source that generates a temperature gradient with
good spherical symmetry.  We assume that the oxide particle is heated
to a surface temperature $T_0$ that is proportional to the laser
power, $T_0 = \alpha P$, which is true if no nonlinear light-matter
interactions occur in the particle.  The particle creates in the fluid
a localized temperature gradient that stabilizes quickly compared to
the timescale of aggregates growth.  Thus, we may assume a steady
state that obeys the time-independent heat equation $\nabla^2 T = 0$,
from which we obtain the spatial profile of the temperature field
$T(r)$, where $r$ is the distance from the center of the oxide
particle:
\begin{equation}
  \label{eq:temp_gradient}
  T(r) = T_{\infty} + \frac {R_0}{r} \Delta T,
\end{equation}
with $\Delta T = T_0 - T_{\infty}$, with $T_\infty$ the bath
temperature far from the titania particle.
PMMA particles in this gradient experience directed motion.  Their
flux $\mathbf j(r)$ through a shell at a position $r$ is
\begin{equation}
  \label{eq:flux}
  \mathbf j(r) = - D(\phi) \, \nabla \phi \, \hat{\mathbf e}_r
  - D(\phi) S_{\text T} \phi \, \nabla T \, \hat{\mathbf e}_r,
\end{equation}
where $D(\phi)$ is the diffusion coefficient at $\phi$, $S_{\text T}$
the Soret coefficient, and $\hat{\mathbf e}_r$ the unit vector normal
to the shell surface.  In this equation, the first term corresponds to
thermal diffusion and the second to the thermophoretic flux.  Solving
the equation $\mathbf j = 0$ gives the steady-state concentration
profile, valid in the fluid phase:
\begin{equation}
  \label{eq:phi_vs_r}
  \phi(r) = \phi_\infty \exp \left( \frac R r \right), 
\end{equation}
with $R = -S_{\text T} \Delta T R_0$, by imposing the
boundary condition $\phi = \phi_\infty$ at $r \rightarrow \infty$.
If there is no metastability in the system, the density of the fluid
phase near the fluid/solid interface should always be close to the
same value, the volume fraction of the phase transition $\phi_{\text
  t}$.  Therefore, in the steady state, the position of the interface
can be obtained using a second boundary condition: $\phi(r_\infty) =
\phi_{\text t}$, where $r_\infty$ is the position of the interface in
the steady state.  We deduce from this condition that the position of
the interface follows:
\begin{equation}
  \label{eq:r_i_formula}
  r_\infty(\phi_\infty, P) = \frac{- S_{\text T} \alpha R_0 P}{\ln
    \frac{\phi_{\text t}}{\phi_\infty}}.
\end{equation}

This equation agrees with the linear dependency of $r_\infty$ on $P$
and the nonlinearity with $\phi_\infty$ observed in the experiments.
It shows that the size diverges when $\phi_\infty \rightarrow \phi_t$,
as minute concentration changes would then lead to the phase
transition.  It can be noted that the only input from thermodynamics
in determining $r_\infty$ is the value of $\phi_{\text t}$.  We fitted
the parameters $- S_{\text T} \alpha R_0$ and $\phi_{\text t}$ to the
data in the inset of \myfig{fig:sizes_graphs}(b) using
\myeq{eq:r_i_formula} and obtained $-S_{\text T} \alpha R_0 =
6.57~\mu$m$\cdot$W$^{-1}$ and $\phi_{\text t} = 0.458$.  The fit
(solid line) shows excellent agreement with the experiments.  The
value found for $\phi_{\text t}$ is very close to the freezing point
of hard spheres at 0.494, which is expected in this system where the
charges of the colloids are
screened~\cite{leocmach13,arai17,royall13,tsurusawa2019direct}.  To
determine $S_{\text T}$, an estimate of $\Delta T$ is required.
Temperatures at such small scales are hard to measure accurately.  A
rough estimate was however obtained by using a sample of titania
particles in a lutidine-water mixture (29.13~\% in mass) that demixes
at a temperature of 33.7~$^\circ$C~\cite{mirzaev06}.  By controlling
the external temperature and measuring the laser power from which
demixing starts when trapping a titania particle, we were able to
estimate the coefficient $\alpha$ to $\alpha \sim 4$~K$\cdot$W$^{-1}$,
from which we deduce $S_{\text T} \sim - 1$~K$^{-1}$ for particles
with $0.40~\mu$m radius.  For particles of radius $1.2~\mu$m, on the
other hand, it was harder to obtain reliable exponential convergence
fits as in \myfig{fig:sizes_graphs}(a): at long times, the size tended
to fluctuate around an average value, because the reorganization of
internal structures occurred.  Movie S6 and
\myfig{fig:big_parts_exp_conv} show the aggregate size growth in a
case where these fluctuations were small, with $\phi_\infty = 0.10$
and $P = 1.5$~W.  We found here $r_\infty = 17~\mu$m and $\tau =
680$~s (see Appendix~\ref{app:1_2_parts}).  Compared to the value of
$r_\infty =6$~$\mu$m of the linear fit in the same conditions in
\myfig{fig:sizes_graphs}(b), for which particles were three times
smaller, we find that $S_{\text T}$ is proportional to $a$.  This is
consistent with a size-independent thermophoretic
mobility~\cite{braibanti08,wurger09}, $D_{\text T} = D S_{\text T}$,
assuming that $D(\phi) \approx k_{\text B} T / (6 \pi \eta a)$ with
$k_{\text B}$ the Boltzmann constant and $\eta$ the solvent viscosity.

In the model above, we assumed that the thermal conductivity of the
mixture was independent of the concentration of PMMA particles.  Since
these have a higher conductivity than the solvent, their concentration
around the central colloid changes the temperature gradient.  We
present here briefly how this changes the equations governing the
steady-state distribution of the particles.  Full justification of the
approximations and details on the calculation are provided in
Appendix~\ref{app:var_tc_model}.  Since the thermal conductivities of
the particles, $k_{\text p}$, and of the solvent, $k_{\text{sol}}$,
only differ moderately, it is appropriate to model the overall thermal
conductivity $k$ of the suspension as the linear combination $k = \phi
k_{\text p} + (1 - \phi) k_{\text{sol}}$ (see
Appendix~\ref{app:var_tc_model}).  This makes $\nabla T$ dependent on
the $\phi(r)$ distribution in \myeq{eq:flux}.  As it is shown in
Appendix~\ref{app:var_tc_model}, this changes the differential
equation satisfied by $\phi(r)$ and its solution, when applying
appropriate boundary conditions. Instead of \myeq{eq:phi_vs_r}, we
obtain
\begin{equation}
  \phi(r) = \frac{1}{\tilde k} W_0 \left( \tilde k \phi_\infty e^{\tilde k
    \phi_\infty + R/r} \right), 
\end{equation}
where $\tilde k = k_{\text p} / k_{\text{sol}} - 1 > 0$ and with $W_0$
the principal branch of the Lambert function.  The position of the
interface in the steady state, $r_\infty$, then becomes
\begin{equation}
  \label{eq:modified_r_inf}
  r_\infty(\phi_\infty, P) = \frac{-S_{\text T} R_0 \alpha P}{\operatorname{ln}
    \frac{\phi_{\text t}}{\phi_\infty} + \tilde k (\phi_t -
    \phi_\infty)},
\end{equation}
instead of \myeq{eq:r_i_formula}.  While the difference in thermal
conductivities is not negligible, using this more accurate model to
fit $- S_{\text T} \alpha R_0$ and $\phi_{\text t}$ gives very similar
values (see Appendix~\ref{app:var_tc_model} for a comparison of the
two models).

\subsection{Dynamics of growth}

Understanding the dynamics of the growth of aggregates is more complex
than determining their final sizes.  For laser powers sufficiently
high to form clusters, the size starts to increase as soon as the
power is raised.  There is, therefore, no sign of existence of
critical nuclei, and the potential gradient is, in that case, strong
enough in the vicinity of the central particle to overcome the energy
barrier for nucleation.  Using slightly lower laser powers to try to
go below this barrier lead to no clustering at all.  These two
observations suggest that critical nuclei must have a size smaller
than the typical range of the decay of the potential, i.e. smaller
than $R_0$ (see \myeq{eq:temp_gradient}).  Even with a negligible
effect of nucleation, the characteristic time $\tau$ is difficult to
obtain analytically as solving the concentration variation $\partial
\phi / \partial t + \nabla \cdot \mathbf j = 0$ requires to include
the advancement of the interface over time and how the flux is written
at this location.

Concerning the advancement, it may be limited by either the reaction
rate to grow crystallites in a supersaturated layer, or by the supply
of PMMA colloids from thermophoresis and diffusion~\cite{gasser09}.
We noted that in the fluid phase close to the interface, the
concentration is high only in a layer of a few colloid diameters.
Therefore, we can rule out that we are in a reaction rate-limited
regime, as there is no accumulation of dense fluid.  In a
diffusion-limited scenario without thermophoresis, a depletion of
particles in the fluid near the growing crystallites is
expected~\cite{gasser09}.  However, we did not observe any depletion.
Furthermore, \myfig{fig:sizes_graphs}(a) shows higher growth rates in
short times for larger temperature gradients.  Hence, we could denote
the phenomenon as ``thermophoresis-limited'' rather than
diffusion-limited.  This denomination should, however, be taken with
care, as there is a complex coupling between diffusion and
thermophoresis in setting the concentration field.  This illustrates
that, in out-of-equilibrium systems, the dynamics of phase transitions
cannot be simply classified as either reaction rate- or
diffusion-limited.

\subsection{Internal structure of aggregates}

\begin{figure}
  \centering
    \includegraphics{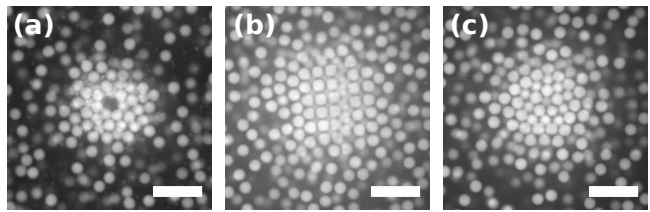}
    \caption{\label{fig:internal_structures} Internal structure of
      condensed clusters (colloids radius: 1.2~$\mu$m, scale bar:
      $10~\mu$m).  Depending on the experimental conditions, amorphous
      (a), polycrystalline with at least fcc crystallites (b) and
      structures with icosahedral order (c) can be obtained.  In (b,c)
      the colloids solution was filtered to remove particles much
      smaller than the 1.2~$\mu$m radius particles that are visible in
      (a).  The high polydispersity in (a) is likely to have prevented
      crystallization.  Furthermore, the density match was better in
      (b) (21.6~\% \emph{cis}-decalin/bromocyclohexane) than in (a)
      and (c) (20~\%), which can affect the kinetics of particles
      addition to the clusters.}
\end{figure}

While we have mainly studied a case where we obtained polycrystalline
structures, \myfig{fig:internal_structures} shows, using 1.2~$\mu$m
radius colloids, that different types of clusters can be obtained, and
we have seen amorphous states (a), crystalline states (b) and
structures with icosahedral order (c).  We do not have a full
quantitative characterization of these systems, but we may explain
these phenomena on a qualitative level.  First, frustration against
crystallization can emerge from various sources~\cite{Tanaka_review}.
The image in \myfig{fig:internal_structures}(a) shows that, in that
experiment, very small particles were also present so that this
polydispersity of the system may be the reason for the amorphous
formation.  Even though the overall size polydispersity is small,
there is a possibility that a small portion of particles with very
different sizes may have a significant impact on the ordering.  Next,
the formation of clusters with icosahedral symmetry like in (c) may
emerge from self-induced spherical confinement. It was
shown~\cite{de2015entropy} that hard spheres compressed under
spherical confinement spontaneously crystallize into icosahedral
clusters that are entropically favored over the bulk fcc crystal
structure.  In our experiments, confinement intervenes in two ways
that are somewhat different from Ref.~\cite{de2015entropy}, but still
with spherical symmetry: through the temperature gradient that induces
a soft and long-range component, and through the central, spherical
impurity that hinders PMMA particles from reaching positions of radius
lower than $R_0 + a$ and that represents an ``inverse spherical
confinement'' with a forbidden void.  Why no long-range icosahedral
order exists in (b) might then be due to different kinetics of
particle addition to the cluster, to slightly different interparticle
interactions as the density and refractive index match was more
accurate in (b) than in (c). Since the first layers of the icosahedral
structure form very early and tend to remain stable as an aggregate
grows (see movie S7, 100x play speed), the symmetry of particle
arrangement initially formed around the titanium particle is likely to
have a substantial impact on the later ordering.  These examples show
very rich possibilities of ordering behaviors in localized
inhomogeneous fields.  We expect that, by manipulating the shape of
the temperature gradient and/or the steric constraints due to the the
metal oxide particle by changing its shape, the symmetry selection of
the clusters of particles should be affected~\cite{hermes11}.

\section{Conclusion}

To summarize, we studied how colloids undergo phase transitions or
transitions to various states under a temperature gradient formed by a
point-like heat source.  We find that thermophoresis leads to the
densification of colloids around the point, resulting in local
crystallization, glass transition, and the formation of structures
with icosahedral order that could be a precursor to a quasicrystal.

This setup provides the advantage of exploring the state growth in
unconventional potentials that have sharp variations in the range of a
few particle diameters but that are still smoother than hard-wall
boundary conditions. Fluctuations in the shape of the interface from a
perfect sphere may, for example, be affected by this. The unique
spatial profile of the potential introduces not only the radial
symmetry but also strong spatial inhomogeneity to the
externally-induced interparticle interaction. Moreover, growth can be
observed by starting from highly non-equilibrium conditions, and our
study indicates that this leads to structures built layer by layer
(although we occasionally saw particle rearrangements in
clusters). This is in stark contrast with the typical first-order
transition in a thermal system, which takes place by overcoming the
free-energy barrier. Finally, the central particle acting as a heat
source can act as a source of geometrical frustration, depending upon
its size and shape. These together with the various structures we
could obtain suggests a possible role of the structure of the
initially formed particle layer and hence the size and shape of the
central impurity on the determination of the ordered structures, which
could be studied further.

\section*{Acknowledgements}

We thank Isaac Theurkauff for his help on the optical tweezers setup
and Taiki Yanagishima for his knowledge on colloids synthesis as well
as useful discussions with Yoav Tsori.  We thank the Institut de
Biologie de Valrose and the PRISM platform for the use of a confocal
microscope to test the samples with a heated platinum wire.  This
study was partially supported by Grants-in-Aid for Specially Promoted
Research (Grant No. JP25000002) and Scientific Research (A)
(JP18H03675) from the Japan Society for the Promotion of Science
(JSPS). N.B. is grateful to JSPS for a JSPS Postdoctoral Fellowship.

\appendix

\section{\label{app:size_meas} Measurement of the size of
  aggregates}

\begin{figure}
  \centering
  \includegraphics[width=8cm]{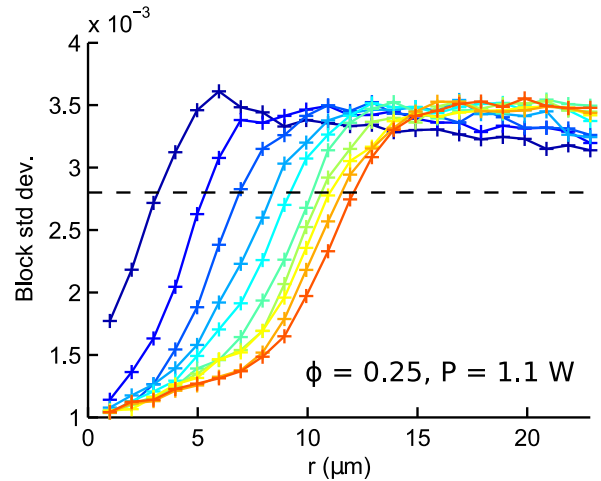}
  \caption{\label{fig:r_from_threshold} Radial evolution of the
    normalized block standard deviation.  The curves represent
    different times (from blue to red: 0 to 1710~s by 190~s steps) and
    show a sharp increase as the solid/fluid interface is passed.
    Applying a threshold (dashed line) gives an estimate of the
    position of the interface.}
\end{figure}

Here we explain how to determine the size of the solid part of the
aggregates from movies consisting of time series of images taken at
1~s intervals.  Because of the small size of the PMMA particles, the
high density of the aggregates, and the limitation of the recordings
to a two-dimensional plane, particle tracking could not be performed
accurately.  We instead measure a locally averaged quantity related to
the mobility of the particles.  The steps are illustrated in
\myfig[s]{fig:blockstd_method} and \ref{fig:r_from_threshold}.

\begin{figure*}
  \centering
  \includegraphics[width=14cm]{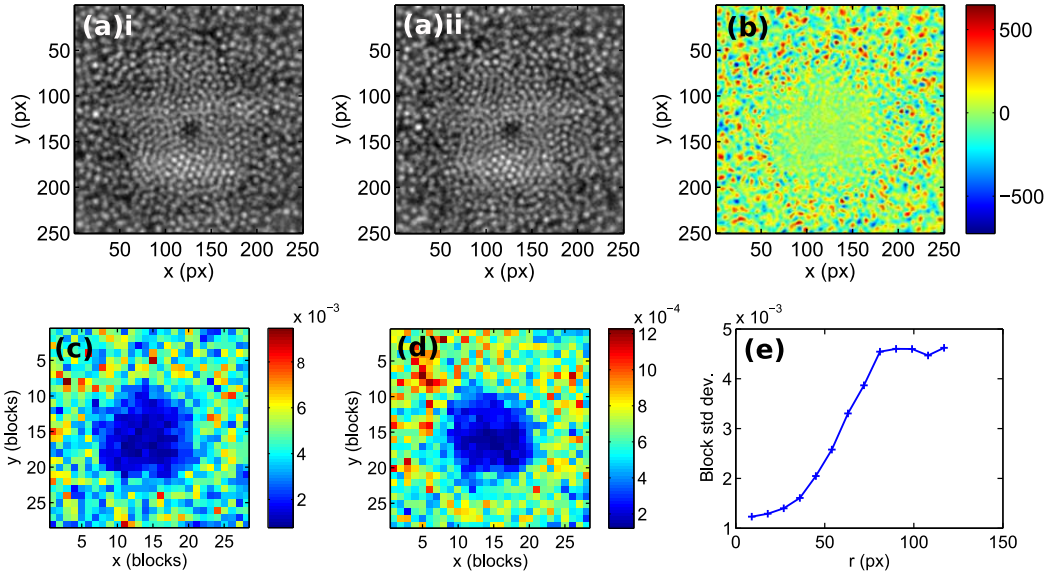}
  \caption{\label{fig:blockstd_method} Extracting a local quantity
    related to the mobility of the particles.  (a) Two frames (Fourier
    filtered) and separated by 10~s.  (b) Intensity difference between
    the two frames in (a).  (c) Block standard deviation map.  (d)
    Block standard deviation map averaged over several time windows.
    (e) Angle-averaged radial variation of the block standard
    deviation.}
\end{figure*}

After applying a band-pass Fourier filter to the images
(\myfig{fig:blockstd_method}(a)), we measure the subtraction of frames
separated by a time delta of 10~s (\myfig{fig:blockstd_method}(b)), a
duration during which free particles typically diffuse by more than
one diameter ($\sqrt{2 D \Delta t} = 2.0~\mu$m, with $D = k_B T / (6
\pi \eta a)$, taking $\eta = 2.6$~mPa$\cdot$s).  The frame difference
maps are then split into square areas called ``blocks'', in a grid
with the size of each block set to 9 pixels (1.0~$\mu$m), of the order
of a PMMA particle diameter.  For each block, we then calculate the
standard deviation of the signal within the block and normalize it by
the average intensity (averaged over the block pixels and the two
frames considered) (\myfig{fig:blockstd_method}(c)).  The map in (c)
shows a well defined central region with very small motion of the
colloids whose size matches the size of the ordered part of the
aggregate seen in (a) and (b).  Since a block can typically contain at
most one particle, the block signal can still fluctuate significantly
for colloids in the gas phase, especially for low volume fractions.
Hence, a further average is performed by calculating and averaging
normalized block standard deviation maps at different times $t$, $t +
\Delta t$, \dots, $t + n_{\text{avg}} \Delta t$ with $n_{\text{avg}} =
8$.  Such an averaged map is displayed in
\myfig{fig:blockstd_method}(d).  To extract an aggregate size, we
reduce the map to a signal depending on the distance $r$ to the center
of the trapped titania particle, averaging over all angles
(\myfig{fig:blockstd_method}(e)).  This signal shows a sharp increase
in the mobility of the particles when traversing the interface between
the two phases.  Finally, we use a threshold of 2.8 to obtain the
aggregate size.  The value was chosen to match interface positions
determined manually on a random set of movies.

An advantage of this method is that regions at different
concentrations, but with particles moving freely, lead to the same
value of the block standard deviation.  The main caveat is that while
the block standard deviation value decreases monotonically with
increasing confinement, there is a nonlinearity that arises from the
intensity profile of a single particle as it affects how the map of
frames difference evolves with the distance a particle has moved.
This nonlinearity is not meaningful as it has no relation with the
fluctuations of the particle positions.

\section{\label{app:1_2_parts} Aggregate size for particles of
  1.2~$\boldsymbol \mu$m radius}

\begin{figure}
  \centering
  \includegraphics[width=8cm]{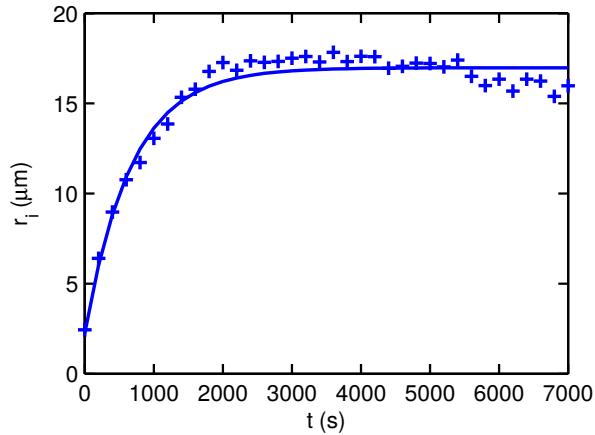}
  \caption{\label{fig:big_parts_exp_conv} Aggregate size $r_{\text i}$
    for $a = 1.2~\mu$m, $\phi_\infty = 0.10$ and $P = 1.5$~W (markers:
    experiment, solid line: fit).}
\end{figure}

Movie S6 (100x play speed) shows the growth of an aggregate of PMMA
particles of radius $a = 1.2~\mu$m and for $\phi_\infty = 0.10$ and $P
= 1.5$~W.  While most of our experiments with these particles showed
strong fluctuations of the size around a final value (after the
initial growth), this particular test has limited fluctuations at long
times, as can be seen in \myfig{fig:big_parts_exp_conv}.  Fitting a
characteristic final size and time of growth gives $r_\infty =
17~\mu$m and $\tau = 680$~s.

\section{\label{app:var_tc_model} Model with concentration-dependent
  thermal conductivity}

In our system, solvent and PMMA particles have different thermal
conductivities.  This should modify the temperature gradient,
depending on the concentration field, in a complex manner due to their
coupling.  We present here a model that uses a simple relation between
thermal conductivity and concentration and derives new formulas for
the concentration field and the size of the aggregates.

\begin{figure}
  \centering
  \includegraphics[width=8cm]{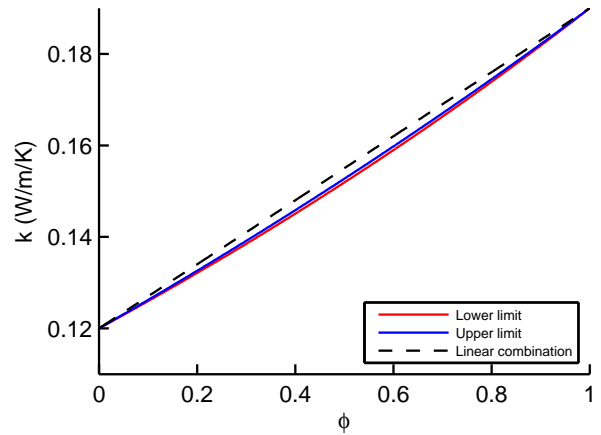}
  \caption{\label{fig:maxwell_model} Thermal conductivity of a
    colloidal suspension, according to the Maxwell mean-field theory.
    For PMMA particles density-matched in a
    \emph{cis}-decalin/bromocyclohexane mixture, the two limits of the
    theory are, for all volume fractions, close to a simple linear
    combination of the conductivities of the colloids and the
    solvent.}
\end{figure}

The Maxwell mean-field theory, initially applied to electric
transport~\cite{maxwell73}, is commonly used to estimate the thermal
conductivity of colloidal suspensions, especially for nanoparticles.
The theory provides formulas for the overall thermal conductivity of
the suspension in two limit cases called ``upper'' and ``lower
limits'' that correspond to different orderings of the particles.
Most experimental data lie between the two limits~\cite{lotfizadeh14}.
These are:
\begin{equation}
  \frac k {k_{\text{sol}}} = 1 + \frac{3 \phi (k_{\text p} - k_{\text{sol}})}{3
    k_{\text{sol}} + (1 - \phi) (k_{\text p} - k_{\text{sol}})}
\end{equation}
for the lower limit, and
\begin{equation}
  \frac k {k_{\text{sol}}} = \frac{k_{\text p}}{k_{\text{sol}}} \left[
    1 - \frac{3 (1 - \phi) (k_{\text p} - k_{\text{sol}})}{3
      k_{\text p} - \phi (k_{\text p} - k_{\text{sol}})} \right]
\end{equation}
for the upper limit with $k$, $k_{\text p}$ and $k_{\text{sol}}$ the
thermal conductivities of the suspension, the PMMA particles and the
solvent, respectively.  Typical values for our system are $k_{\text p}
= 0.19$~\cite{assael05} and $k_{\text{sol}} =
0.12$~W/m/K~\cite{kim16}.  \myfig{fig:maxwell_model} shows that for
this moderate discrepancy between the two thermal conductivities, the
overall conductivity of the suspension given by the Maxwell theory can
be approximated to the linear combination
\begin{equation}
  \label{eq:conductivity_linear}
  k = \phi k_{\text p} + (1 - \phi) k_{\text{sol}} \, .
\end{equation}
To simplify the calculation below, we will therefore model the
conductivity in our system with \myeq{eq:conductivity_linear}.

We write $P_{\text h}$ the power of heat dissipated by the central
colloid.  In the steady-state, the heat flow through any sphere at
radius $r>R_0$ should equate $P_{\text h}$.  In a spherically
symmetric configuration, the heat flux is $\mathbf q = -k \frac{\text
  d T}{\text d r} {\hat{\textbf e}}_r$, so that
\begin{equation}
  \label{eq:heat_flux}
  \frac{P_{\text h}}{\pi r^2} = -k \frac{\text d T}{\text d r}
\end{equation}
where $k$ now depends on the concentration of colloids according to
\myeq{eq:conductivity_linear}.
As in the main text, the motion of the PMMA particles is governed in
the steady state by the balance of flux due to diffusion and the
thermophoretic flux that are both in the $\hat{\textbf e}_r$
direction:
\begin{equation}
  \label{eq:part_flux_balance}
  0 = -D \frac{\text d \phi}{\text d r} - D S_{\text T} \phi \frac{\text d
  T}{\text d r} \, .
\end{equation}
If there is a phase transition with a solid aggregate for $r <
r_\infty$, this equation is then only valid for $r > r_\infty$.  In
this range, \myeq[s]{eq:conductivity_linear}, (\ref{eq:heat_flux}) and
(\ref{eq:part_flux_balance}) give the following differential equation
in $\phi(r)$:
\begin{equation}
  \label{eq:diff_eq_0}
  r^2 \left(1 + \tilde k \phi \right) \frac{\text d \phi}{\text d r} -
  \frac{S_{\text T} P_{\text h}}{\pi k_{\text {sol}}} \phi = 0
\end{equation}
where $\tilde k = k_{\text p} / k_{\text{sol}} - 1$.  It is
appropriate to express $P_{\text h}$ as function of quantities
measured experimentally.  Since $P_{\text h}$ is not dependent on how
the heat propagates in the fluid, we obtain it in our work from the
measurement (see the main text) of the temperature difference between
the surface of the colloids and the temperature far from the center,
$\Delta T = T_0 - T_\infty$, in a fluid without colloids.  $P_{\text
  h}$ is then deduced by integrating $\frac{P_{\text h}}{\pi r^2} =
-k_{\text{sol}} \frac{\text d T}{\text d r}$ between $R_0$ and $r
\rightarrow \infty$.  We obtain:
\begin{equation}
  P_{\text h} = \pi k_{\text{sol}} R_0 \Delta T
\end{equation}
and \myeq{eq:diff_eq_0} becomes:
\begin{equation}
  \label{eq:diff_eq_1}
  r^2 \left(1 + \tilde k \phi \right) \frac{\text d \phi}{\text d r} +
  R \phi = 0
\end{equation}
with $R = -S_{\text T} \Delta T R_0$.  The solutions to this equation
are expressed in terms of the Lambert $W$ function, which only has two
branches in real space.  Imposing the constraint that $\tilde k > 0$,
the relevant branch is the principal branch $W_0$ and we deduce
\begin{equation}
  \phi(r) = \frac{1}{\tilde k} W_0 \left( \tilde k e^{C +  R/r} \right)
\end{equation}
with $C$ a constant.  The constant is obtained by applying the
boundary condition that the concentration tends to $\phi_{\infty}$
when $r \rightarrow \infty$.  Using the property that $z = W_0(z e^z)$
for any $z$ value, we get $C = \operatorname{ln} \phi_\infty + \tilde
k \phi_\infty$ and the concentration field:
\begin{equation}
  \phi(r) = \frac{1}{\tilde k} W_0 \left( \tilde k \phi_\infty e^{\tilde k
      \phi_\infty + R/r} \right) \, .
\end{equation}
For $\tilde k \rightarrow 0$, we recover the concentration field for
constant thermal conductivity (Eq.~(3) in the main text).

\begin{figure}
  \centering
  \includegraphics[width=8cm]{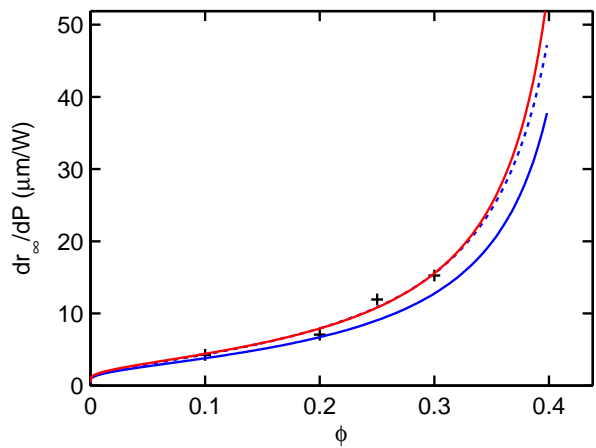}
  \caption{\label{fig:modified_dp_over_dr} Comparison of the estimates
    of aggregate sizes using the models with constant thermal
    conductivity, Eq.~(4) (dashed blue curve; the same as in the main
    text) and $\phi$-dependent conductivity, \myeq{eq:modified_r_inf}
    (solid blue curve) with the same values of the fitting parameters,
    $\phi_\infty$ and $-S_{\text T} R_0 \alpha$.  Then, the red solid
    curve is the fit of $\phi$-dependent conductivity,
    \myeq{eq:modified_r_inf}, with new parameters.}
\end{figure}

Following the same reasoning as in the case of constant thermal
conductivity, and setting $r$ to the steady-state interface position
$r_\infty$, for which $\phi(r_\infty) = \phi_{\text t}$, we find the
position of the interface at
\begin{equation}
  \label{eq:modified_r_inf}
  r_\infty = \frac{-S_{\text T} R_0 \alpha P}{\operatorname{ln}
    \frac{\phi_{\text t}}{\phi_\infty} + \tilde k (\phi_t -
    \phi_\infty)} \, .
\end{equation}
\myfig{fig:modified_dp_over_dr} compares this model (solid blue line)
with variable thermal conductivity to the simpler model in the main
text (dashed line), taking for $-S_{\text T} R_0 \alpha$ and $\phi_t$
the values obtained from fitting the simplified model.  The two models
only match at low volume fractions and differ by a factor $1 + \tilde
k \phi_t$ when approaching $\phi_t$.  In the moderate range of volume
fractions we have explored (up to $\phi_\infty = 30$~\%), the maximum
discrepancy between the two models is about 20~\%.  Fitting the
experiments to \myeq{eq:modified_r_inf} (red curve) gives similar
parameters as for the model with constant conductivity: $-S_{\text T}
R_0 \alpha = 7.47~\mu$m/W and $\phi_t = 0.445$.  The discrepancy
between the two models appears to affect mainly the $-S_{\text T} R_0
\alpha$ parameter.

\end{document}